\documentclass[twocolumn,aps,prl,showpacs]{revtex4-1}
\usepackage{graphicx}
\usepackage{float}
\usepackage{amsmath}
\usepackage{amsfonts}
\usepackage{hyperref}
\usepackage{fullpage}
\newcommand{\mathsym}[1]{{}}

\usepackage{subfig}

\begin{document}
\title{Boundary Conditions for Electron Flow in Graphene in the Hydrodynamic Regime}
\author{Glenn Wagner$^{1,2}$}
\affiliation{$^1$Physics of Complex Systems, Weizmann Institute of Science, Rehovot 76100 Israel $^2$Department of Physics, University of Oxford, Oxford OX1 3PU United Kingdom}

\date{\today}
\begin{abstract}
Graphene has generated a lot of research interest due to its special properties, which include a hydrodynamic regime. It is not yet clear however which boundary condition such a hydrodynamic current flow satisfies. The aim of this paper is to investigate the effect of different boundary conditions on the potential in an infinite strip of graphene, in which the electrons can be treated hydrodynamically. The boundary conditions on the current range continuously from no-slip to a free boundary. We analyse the situation for two different orientations of the source and sink, inspired by recent papers. We discuss which geometry is better suited for identifying the hydrodynamic regime and experimentally determining the boundary conditions.
\end{abstract}

\maketitle

\section{Introduction}

Correct boundary conditions are crucial but non-trivial for solving many problems in physics. One such problem is the flow of a fluid in a bounded region. For a classical fluid it is well-known that at the boundary between a fluid and a solid, the no-slip condition accurately describes the behaviour of the fluid \cite{FluidMech}. The strong forces between the fluid and the solid make the fluid stick to the boundary, in all but very unusual situations. Despite being well-accepted nowadays, this condition was a matter of controversy for many years after it was introduced by George Stokes in the mid 19th century\cite{NoSlip}. On the other hand, at the boundary between a liquid and a gas, the free boundary condition has to be applied to the liquid.

More recently, work has been done on the hydrodynamic regime in graphene, in which the electrons satisfy equations analogous to the fluid mechanical Navier-Stokes equation for an incompressible fluid \cite{Geim}. The electrons in graphene can behave as a Fermi liquid and have very high mobilities. To observe this regime we must have a very pure sample in a specific temperature range: The temperature must be low enough to reduce scattering of the electrons off the lattice, but high enough such that the mean free path of electrons due to collisions with each other is much smaller than the sample size. This can occur at temperatures up to room temperature \cite{Geim}. The question remains: What conditions should be applied at the boundary? Unlike in the fluidic case it is not clear whether to apply the no-slip condition here. 

Two recent papers \cite{Geim}, \cite{FL} have looked for possible experimental signatures of the hydrodynamic regime in graphene. The papers both use the geometry of an infinite strip of width $W$, however different boundary conditions are employed. In \cite{FL} an arrangement is used in which source and sink are on opposite sides of the strip, whereas in \cite{Geim} source and sink are on the same side. In addition \cite{FL} focusses more on the no-slip conditions at the boundary, while \cite{Geim} focusses more on the free boundary condition. In this paper both geometries are investigated and the advantages and disadvantages of each are compared. The criteria are the suitability for identifying the hydrodynamic regime and the correct boundary conditions and for measuring the viscosity. In \cite{FL} it has been suggested that the negative nonlocal resistance can be used as an identifier of the hydrodynamic regime in graphene, since it does not arise for zero viscosity. In this article we discuss the question of how robust this identifier is with respect to the boundary conditions.

We describe the boundary condition by a continuous parameter $l_{b}$, the slipping length, as in \cite{Geim}. The possible regimes range from the no-slip condition $l_{b}=0$ to the free boundary condition $l_{b}\rightarrow\infty$. In this article we propose a method for experimentally determining the value of the slipping length. 

We focus on the potential at the boundary and explain its spatial behaviour and dependence on the relevant parameters of this problem: the slipping length $l_{b}$ and the viscosity, quantified by $D_{\nu}$, the vorticity diffusion length \cite{Geim} or equivalently $\frac{\rho(enW)^{2}}{\eta}$ \cite{FL}. The correspondence of the dimensionless parameters $d=\frac{D_{\nu}}{W}$ and $g=\frac{\rho(enW)^{2}}{\eta}$ is $d=g^{-1/2}$. Here $\eta$ is the dynamic viscosity, $e$ is the electron charge (a negative quantity), $\rho$ is the ohmic resistivity and $n$ is the electron number density. So g tells us the relative importance of viscosity and resistivity.

In \cite{Geim} and \cite{Geim2} it is claimed that the free boundary conditions are most appropriate for their particular experimental set-up, since the Gurzhi effect is not observed. In this case the Gurzhi effect \cite{Gurzhi} describes the increase of the four-point conductivity with temperature. However, it is important to have a method for determining a finite slipping length $l_{b}$, as it is possible that the type of boundary condition depends on the nature of the boundary and the exact experimental conditions or that $l_{b}$ is large but finite, explaining why the Gurzhi effect is not observed. In the absence of conclusive theoretical arguments for the correct boundary conditions taking into account the interaction of the electrons and the boundary, the best that can be done is to experimentally search for signatures of the boundary conditions.

\section{source and sink on same side}
This is the set-up used in the paper by Torre, Tomadin, Geim and Polini \cite{Geim}.
\begin{figure}[H]
\begin{centering}
\includegraphics[width=1\linewidth]{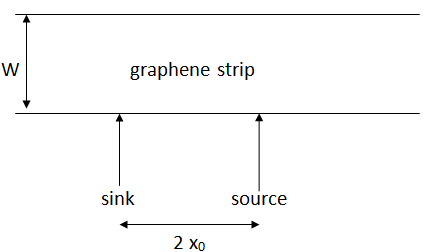}\caption{geometry of the first set-up}

\par\end{centering}

\end{figure}

The origin is at the centre of the strip between source and sink.

\subsection{Set-up}

We use the method of \cite{Geim} to solve the problem of a current
source and sink in an infinite strip of width $W$. The governing equations are the incompressibility (continuity) and Navier-Stokes equations:

\begin{equation}
\nabla\cdot\mathbf{J}=0
\label{incompressibility}
\end{equation}

\begin{equation}
\frac{\sigma_{0}}{e}\nabla\phi+D_{\nu}^{2}\nabla^{2}\mathbf{J}=\mathbf{J}
\label{NavSto} 
\end{equation}
The non-linear term has been neglected, assuming low enough Reynolds number. The first boundary condition is

\begin{equation}
J_{y}(x,y=\pm\frac{W}{2})=J_{\pm}(x)
\label{SourceBC} 
\end{equation}

which defines the position of the source and sink. Thus we fix the current entering or exiting the strip though the contacts and there is no current perpendicular to the boundary away from the contacts. The second boundary condition is

\begin{equation}
\left[\partial_{y}J_{x}+\partial_{x}J_{y}\right]_{y= W/2}=\mp\frac{J_{x}(x,y=\pm\frac{W}{2})}{l_{b}}
\label{SlipBC} 
\end{equation}

The interpretation of this last boundary condition is that the force exerted by the boundary of the fluid is proportional to the tangential velocity.                                                                                                                                                                                                     $l_{b}$ is a phenomenological parameter of dimensions of length which characterizes the slipping, however it does not have a clear physical meaning in itself.

In the first case we use the same set-up as in \cite{Geim}, i.e. a source
and sink both at the bottom edge of the strip at $\pm x_{0}$ . \\

We regularize via the Lorentzian with a characteristic length $l$, so that the source and drain have a finite size, as is physically sensible. Then the first boundary condition becomes

\begin{equation}
J_{+}=0
\label{BC1}
\end{equation}

and

\begin{equation}
J_{-}=\frac{I}{-e}\left(\frac{l}{l^{2}+(x-x_{0})^{2}}-\frac{l}{l^{2}+(x+x_{0})^{2}}\right)
\label{BC2}
\end{equation}

In all the following calculations we put $\frac{l}{W}=1/20$ as in \cite{FL} to allow comparison with their results which use this value. 
After Fourier transforming in $x$, we can put the equations (\ref{incompressibility}) to (\ref{NavSto}) in matrix form and solve them as in \cite{Geim} by diagonalizing the matrix. We find:
\begin{eqnarray*}
\left(\begin{array}{c}
k\hat{J_{x}}\\
k\hat{J_{y}}\\
\partial_{y}\hat{J_{x}}\\
k^{2}(\sigma_{0}/e)\hat{\phi}
\end{array}\right)  = a_{1}\left(\begin{array}{c}
i\\
-1\\
-i\\
1
\end{array}\right)e^{-ky}+a_{2}\left(\begin{array}{c}
i\\
1\\
i\\
1
\end{array}\right)e^{ky}\nonumber \\
 +a_{3}\left(\begin{array}{c}
-\frac{\mid k\mid D_{\nu}}{\sqrt{1+(kD_{\nu})^{2}}}\\
-i(kD_{\nu})^{2}/(1+(kD_{\nu})^{2})\\
1\\
0
\end{array}\right)e^{-ky\sqrt{1+1/(kD_{\nu})^{2}}}\\
 +a_{4}\left(\begin{array}{c}
\frac{\mid k\mid D_{\nu}}{\sqrt{1+(kD_{\nu})^{2}}}\\
-i(kD_{\nu})^{2}/(1+(kD_{\nu})^{2})\\
1\\
0
\end{array}\right)e^{ky\sqrt{1+1/(kD_{\nu})^{2}}}\nonumber 
\label{solution} 
\end{eqnarray*}
where the Fourier transform is denoted by a hat.

The coefficients $a_{i}$ can be determined from the Fourier transform of the boundary conditions (\ref{SourceBC}) and (\ref{SlipBC}). We thus obtain analytical solutions for $\hat{\phi}$, $\hat{J_{x}}$ and $\hat{J_{y}}$. To obtain $\phi$, $J_{x}$ and $J_{y}$ we Fourier transform numerically.

\subsection{Results}

On the following page we plot the potential at the lower edge $\phi (x,y=-W/2)$ in arbitrary units for $\frac{\rho(enW)^{2}}{\eta}=1$ and $\frac{\rho(enW)^{2}}{\eta}=50$. In both cases we plot the result for three different values of the slipping length, corresponding to the no-slip condition, a partial slip and a free boundary. When the potential becomes negative, this corresponds to a negative nonlocal resistance. 

\begin{figure}[H]
    \includegraphics[width=7.0cm]{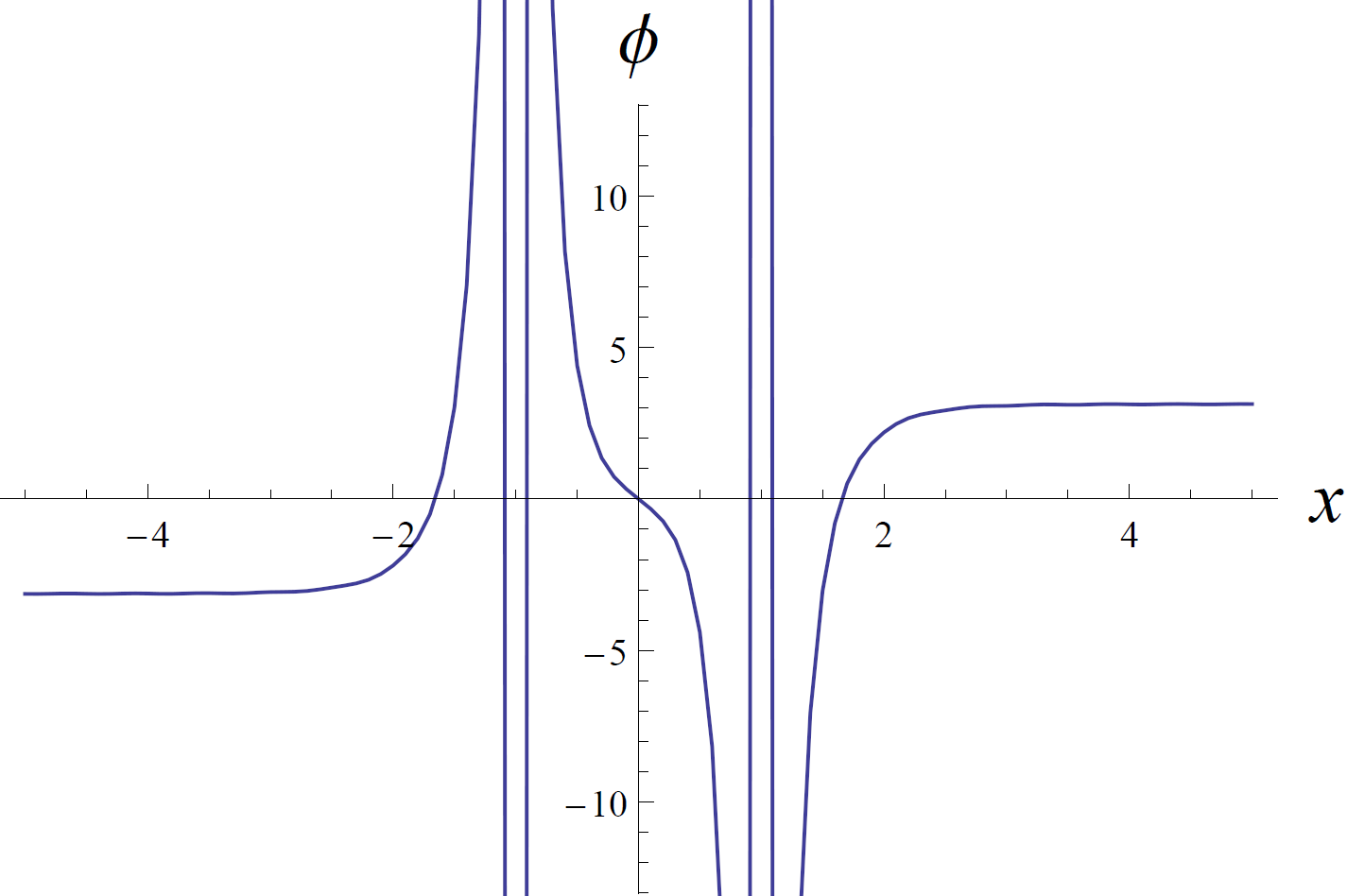}
    \centering
    \caption{$\phi (y=-W/2)$ for free boundary $\frac{l_{b}}{W}=1000$ and $\frac{\rho(enW)^{2}}{\eta}=1$
or $\frac{D_{\nu}}{W}=1$}
    \label{fig:Geim 1000 1}
\end{figure}

For free boundary conditions and $\frac{\rho(enW)^{2}}{\eta}=1$ (Fig. \ref{fig:Geim 1000 1}) we recover the results of the paper \cite{Geim} with the negative resistance. Close to the source/sink the potential has the sign as would be expected without viscosity, this is due to the finite size of the electrode. The ideal delta-function source considered in \cite{Geim} does not have this spike at the sources. 

The potential at infinity also has the same sign as in the ohmic case, the resistive term (the first term in (\ref{NavSto})) dominates over the viscous term (the second term in (\ref{NavSto})) in the regions at large $x$. It is noteworthy, that the potential does not decay to zero, as we go to $\lvert x\lvert\rightarrow\infty$. Instead at positive $x$, for large viscosities, the potential tends to a positive constant from below, there is a flow towards the source due to a vortex. For small viscosities the potential approaches a constant from above, as we approach the ohmic situation. The reason that the potential can tend to a finite constant as $x \rightarrow \pm \infty$ is that we have broken the left-right symmetry. So if we connected the ends of the strip at large $x$, there would be a non-zero, though small,  electron flow due to the potential difference since the boundary conditions define a sense of rotation. 

Moving away slightly from $\pm x_{0}$ the potential changes sign. As explained in \cite{FL} the sign change in the direction away from the origin is due to a vortex appearing. The sign-change closer to the origin can be thought of as due to the free boundary condition $\frac{\partial J_{x}}{\partial y}=0$ at the boundary $y=-W/2$.

\begin{figure}[H]
\begin{centering}
\includegraphics[width=6cm]{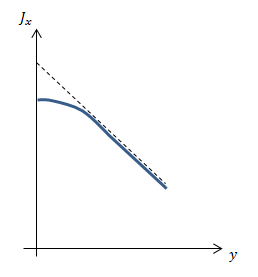}\caption{Illustration of the cause of negative nonlocal resistance}

\par\end{centering}

\end{figure}
Without any boundary and with zero potential the viscous term alone would create a (e.g. linearly) decreasing velocity profile. The boundary condition $\frac{\partial J_{x}}{\partial y}=0$ changes the solution only close to the boundary: It lowers the current at the $y=-W/2$ boundary, as shown above. The voltage thus reverses its sign to impose these boundary conditions. Keeping $\frac{\rho(enW)^{2}}{\eta}=1$ and changing the boundary conditions, an additional sign change in the potential appears when we decrease the slip, such that the sign of the voltage close to the origin is the same as in the Ohmic case. 
This is consistent with the previous argument.

\pagebreak
\subsection{Geometry of \cite{Geim}}
\begin{figure}[H]
    \includegraphics[width=7.0cm]{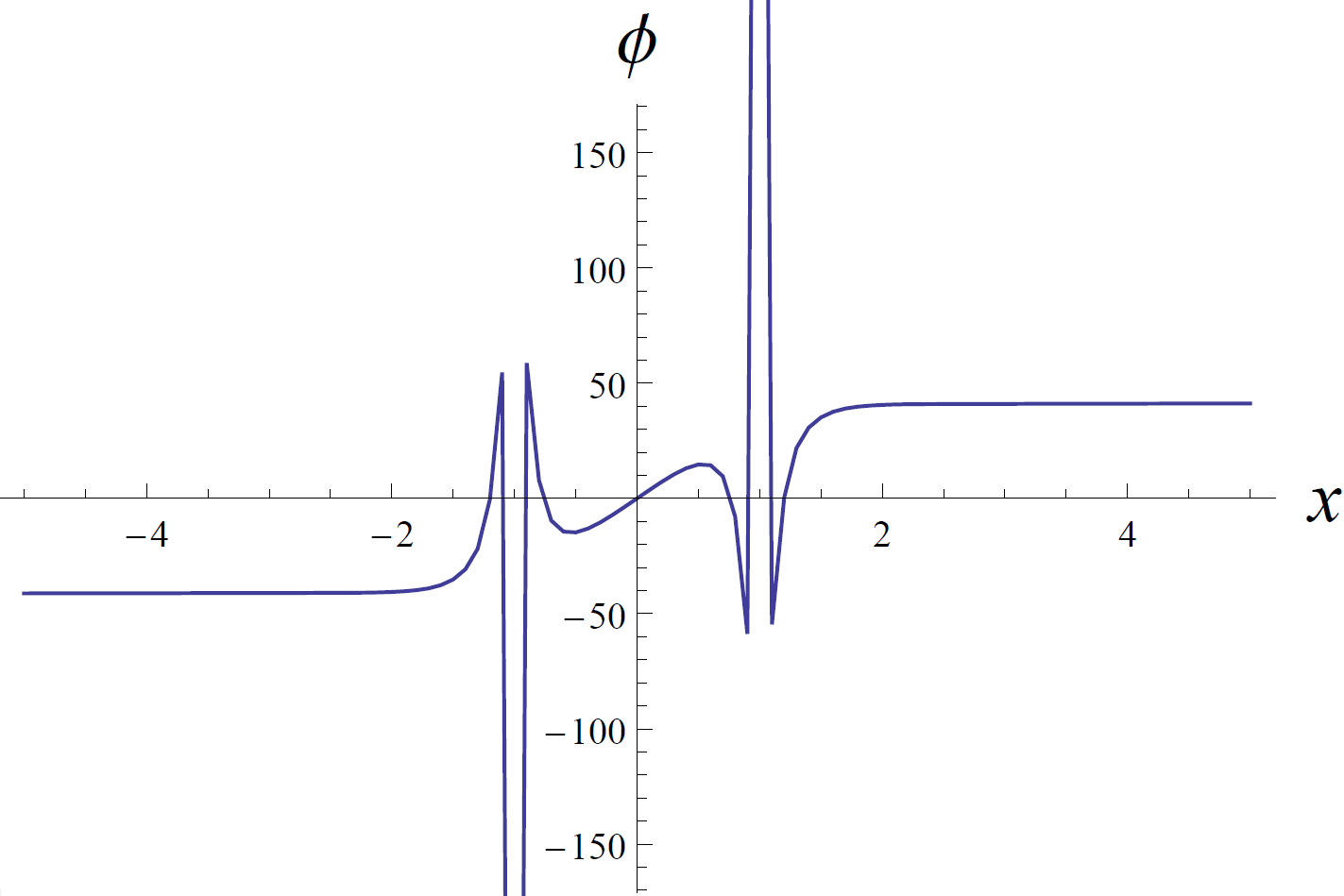}
    \centering
    \caption{$\phi (y=-W/2)$ for no-slip $\frac{l_{b}}{W}=0$ and $\frac{\rho(enW)^{2}}{\eta}=1$ or
$\frac{D_{\nu}}{W}=1$}
    \label{fig:Geim 0 1}
\end{figure}
\begin{figure}[H]
    \includegraphics[width=7.0cm]{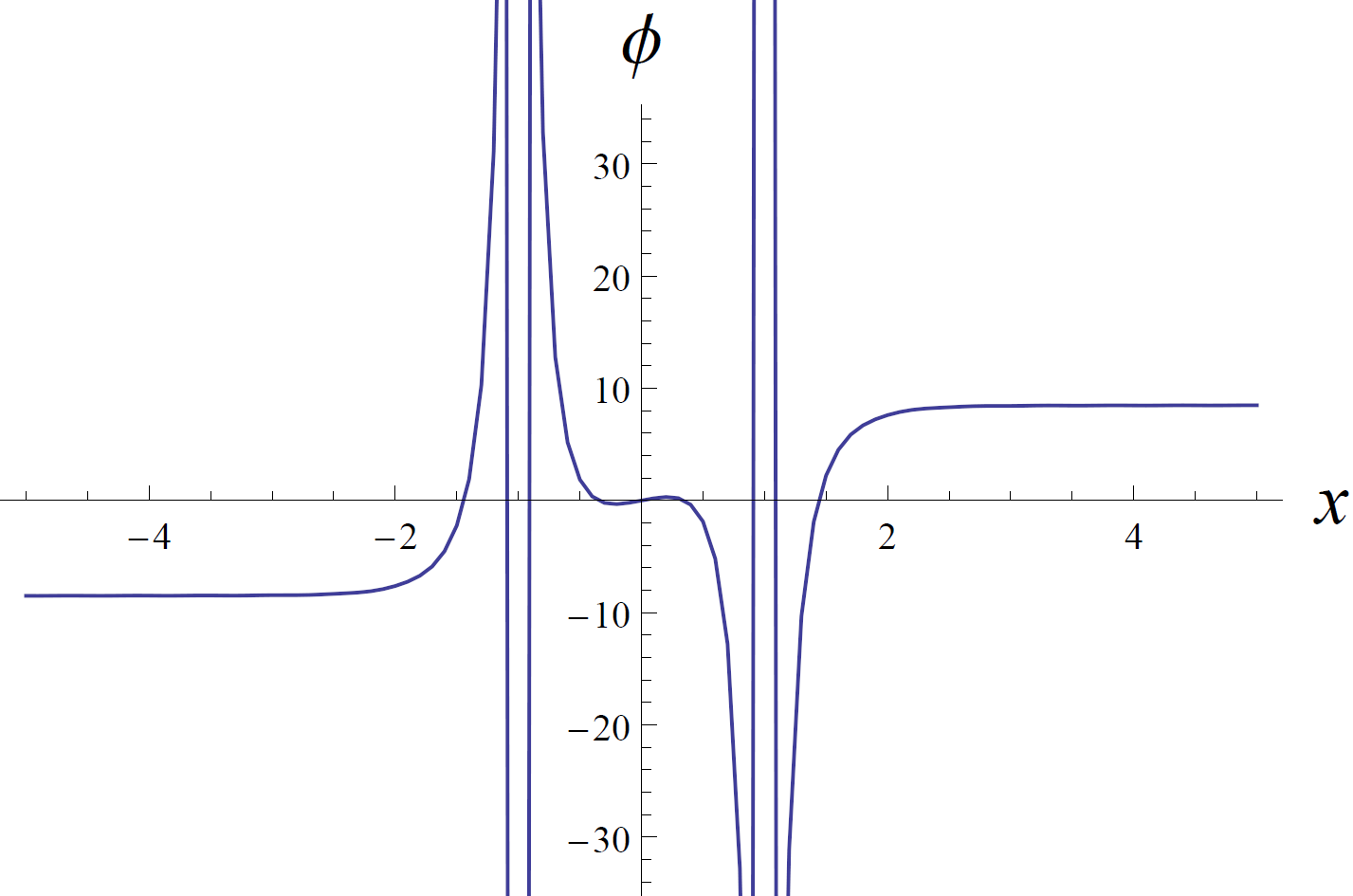}
    \centering
    \caption{$\phi (y=-W/2)$ for partial slip $\frac{l_{b}}{W}=1$ and $\frac{\rho(enW)^{2}}{\eta}=1$
or $\frac{D_{\nu}}{W}=1$}
    \label{fig:Geim 1 1}
\end{figure}
\begin{figure}[H]
    \includegraphics[width=7.5cm]{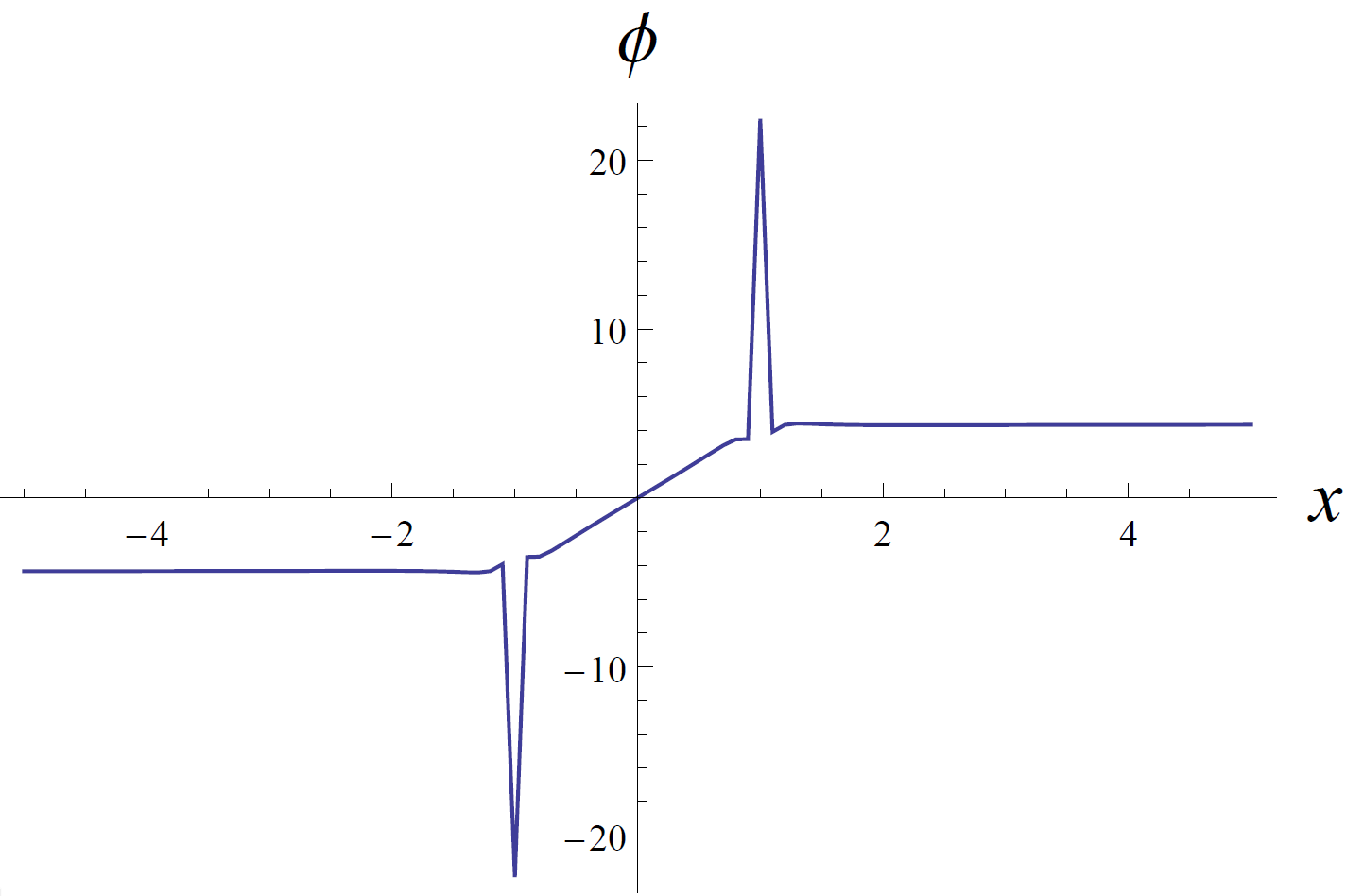}
    \centering
    \caption{$\phi (y=-W/2)$ for no-slip $\frac{l_{b}}{W}=0$ and $\frac{\rho(enW)^{2}}{\eta}=50$
or $\frac{D_{\nu}}{W}=0.14$}
    \label{fig:Geim 0 50}
\end{figure}
\begin{figure}[H]
    \includegraphics[width=7.5cm]{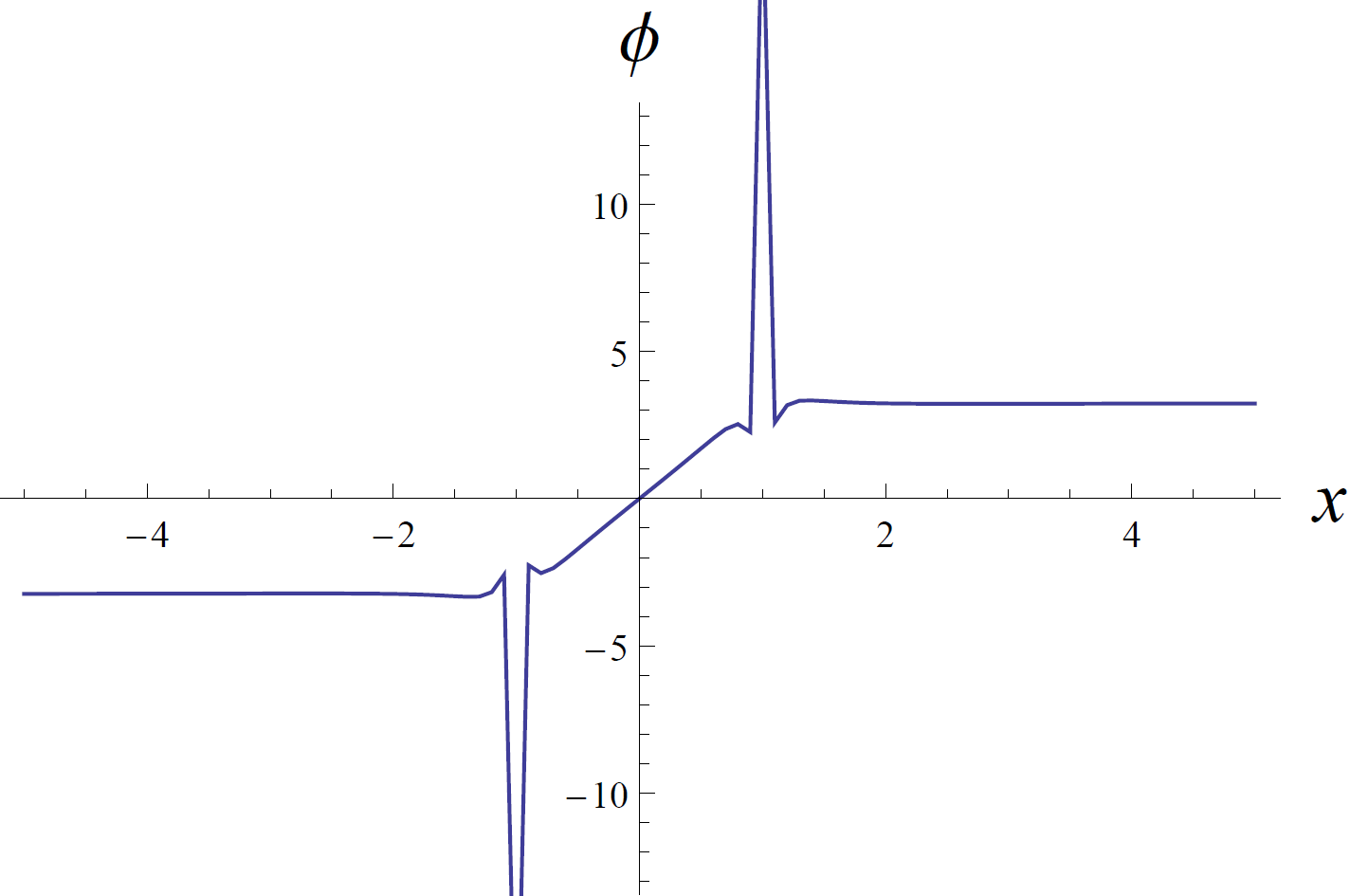}
    \centering
    \caption{$\phi (y=-W/2)$ for partial slip $\frac{l_{b}}{W}=1$and $\frac{\rho(enW)^{2}}{\eta}=50$
or $\frac{D_{\nu}}{W}=0.14$}
    \label{fig:Geim 1 50}
\end{figure}
\begin{figure}[H]
    \includegraphics[width=1\linewidth]{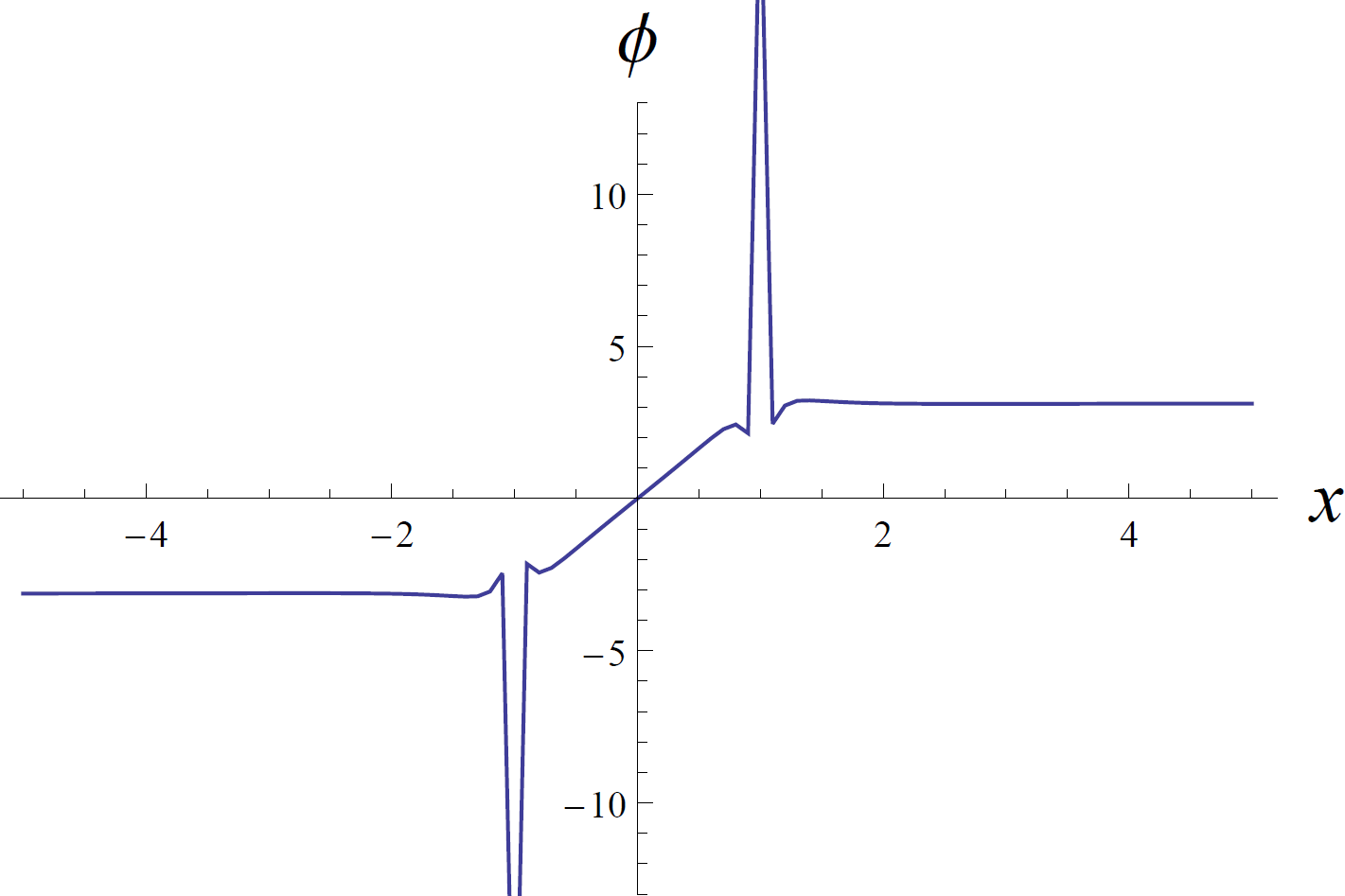}
    \centering
    \caption{$\phi (y=-W/2)$ for free boundary $\frac{l_{b}}{W}=1000$ and $\frac{\rho(enW)^{2}}{\eta}=50$
or $\frac{D_{\nu}}{W}=0.14$}
    \label{fig:Geim 1000 50}
\end{figure}

Decreasing the viscosity such that $\frac{\rho(enW)^{2}}{\eta}=50$ as in Figs. \ref{fig:Geim 0 50} to \ref{fig:Geim 1000 50} we do not observe any negative resistance. We obtain something much more like the Ohmic case and the boundary conditions have little impact on the potential\, the reason being that we are exiting the hydrodynamic regime and the viscous term becomes negligible.

So in this geometry we can identify two regimes for the dependence on the boundary conditions. For low viscosities (as with $D_{\nu}=50$) there is hardly any dependence on the boundary conditions, while for high viscosities (as in $D_{\nu}=1$) the boundary conditions have an important effect. Especially the behaviour of the potential at the origin depends on the boundary conditions, with additional lobes appearing for the no-slip case. This could be used to identify the boundary conditions. 

\begin{figure}
    \includegraphics[width=1\linewidth]{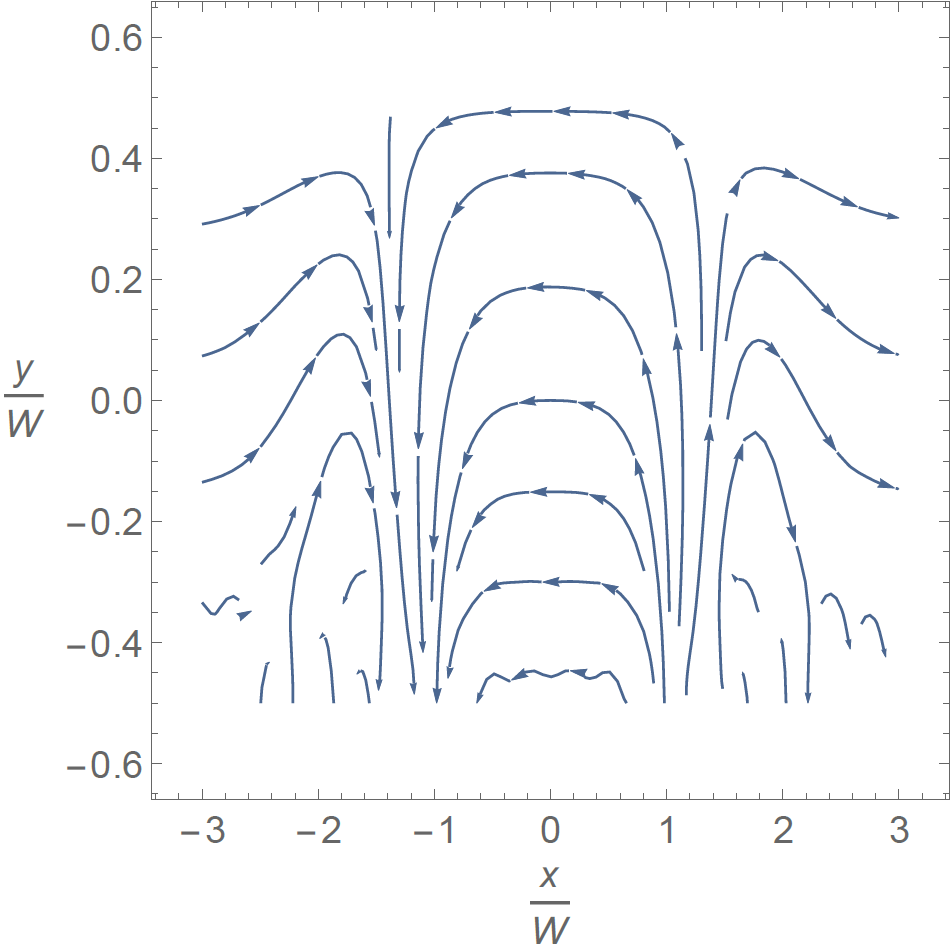}
    \centering
    \caption{Streamlines of the current for partial slip $\frac{l_{b}}{W}=1$ and $\frac{\rho(enW)^{2}}{\eta}=50$ or $\frac{D_{\nu}}{W}=0.14$}
    \label{fig:GeimStream}
\end{figure}

In the streamplot of the current (Fig. \ref{fig:GeimStream}) we can see the vortices appearing near both contact points. This is the cause of the negative nonlocal resistance which exists for these values of the parameters. However the current between source an sink is much stronger than all other currents.

\begin{figure}
    \includegraphics[width=1\linewidth]{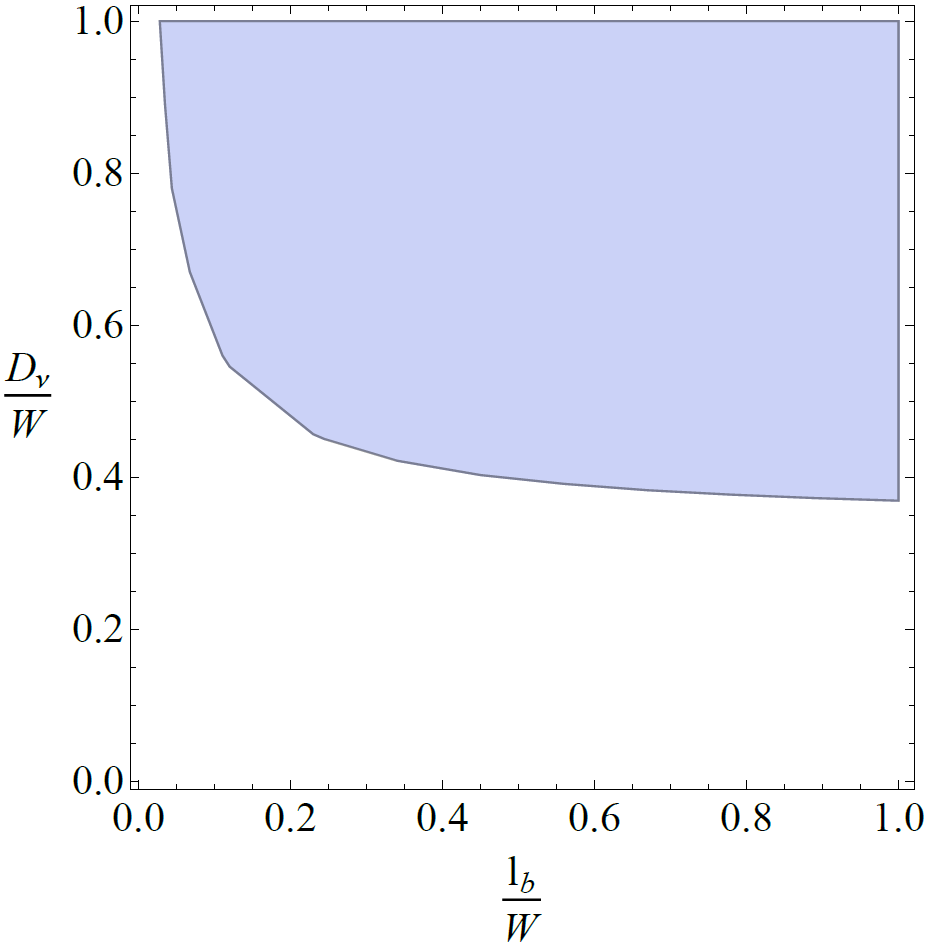}
    \centering
    \caption{coloured region is that region in the $l_{b}$-$D_{\nu}$ parameter space in which $\phi (x,y=-W/2)$ becomes negative (geometry of \cite{Geim})}
    \label{fig:GeimPhase}
\end{figure}

Finally in Fig. \ref{fig:GeimPhase}, we plot the region in the $l_{b}$-$D_{\nu}$ parameter space in which $\phi (x,y=-W/2)$ becomes negative, i.e. the region in which we observe a negative nonlocal resistance. As expected, for large enough viscosity, around $D_{\nu}/W>1$ there is a negative potential regardless of the boundary conditions. The critical value of the viscosity for which negative potential is observed also decreases as we increase $l_{b}$. This agrees with our intuition that the no-slip condition inhibits the negative nonlocal resistance, since vortices at the edge of the strip would create a non-vanishing slip. 
Hence, one possible approach for determining the slipping length $l_{b}$ would be to find the critical value of $D_{\nu}/W$ at which the negative non-local resistance first appears.

\newpage

\section{source and sink on opposite side}
This is the set-up used in the paper by Levitov and Falkovich \cite{FL}.
\begin{figure}[H]
\begin{centering}
\includegraphics[width=1\linewidth]{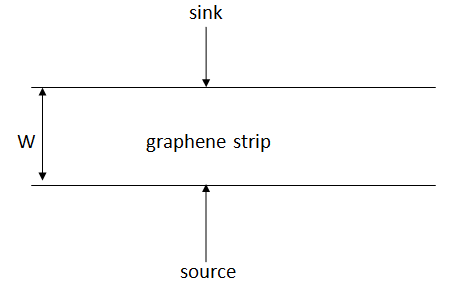}\caption{geometry of the second set-up}

\par\end{centering}

\end{figure}
\subsection{Set-up}

In this case the governing equations are unchanged, but the boundary
conditions (\ref{BC1})-(\ref{BC2}) must be replaced by

\begin{equation}
J_{+}=-\frac{I}{-e}\frac{l}{l^{2}+x^{2}}
\end{equation}

\begin{equation}
J_{-}=\frac{I}{-e}\frac{l}{l^{2}+x^{2}}
\end{equation}

The problem can be solved in the same way as before. Due to additional symmetry, it can be solved by hand, as is done in the appendix.

\subsection{Results}
In the appendix we plot the potential at the lower edge $\phi (x,y=-W/2)$ for $\frac{\rho(enW)^{2}}{\eta}=30$ and $\frac{\rho(enW)^{2}}{\eta}=50$. As before, in both cases we plot the result for three different values of the slipping length, corresponding to the no-slip condition, a partial slip and a free boundary. In the no-slip limit we expect the results to agree with \cite{FL}. Plotting the potential for different values of $D_{\nu}$ at $l_{b}=0$ we find the same general shape of the potential. Indeed the qualitative picture is the same, we observe the same general form of the potential and the shape of the curve changes in the same way when we decrease $\frac{\rho(enW)^{2}}{\eta}$. However the value of $\frac{\rho(enW)^{2}}{\eta}$ for which the curve adopts a particular shape is different. For example the critical value of $\frac{\rho(enW)^{2}}{\eta}$ at which the negative resistance first appears is around 47.5 in our results, whereas in \cite{FL} the corresponding value is 120. So our value of $\frac{\rho(enW)^{2}}{\eta}$ is around 2.5 times as large as theirs for corresponding shapes of the potential. This discrepancy is as yet unaccounted for. We observe that if we apply the no-slip boundary condition to the problem, we do not necessarily find any negative resistance. We also observe the potential to decay to zero at large $x$ as the argument applied to the previous geometry does not hold here.

This set-up allows for the experimental determination of the value of the slipping length, which is a phenomenological parameter. A characteristic feature of the graphs of the voltage in this geometry are the minima at $\pm x_{\mathrm{min}}$, which can occur at a positive or negative potential. The location of the minimum depends on both $D_{\nu}$ and $l_{b}$, so if the viscosity is known, this method can be used to compute the slipping length. Indeed, \cite{PerfFluid} find an expression for the viscosity of the electrons in graphene. The  x-value of the minimum of the potential is not always a monotonic function of $l_{b}$, but in the case where it is not, a general plot of the potential along the boundary should allow the different possible values of $l_{b}$ to be distinguished.

\begin{figure}
    \includegraphics[width=1\linewidth]{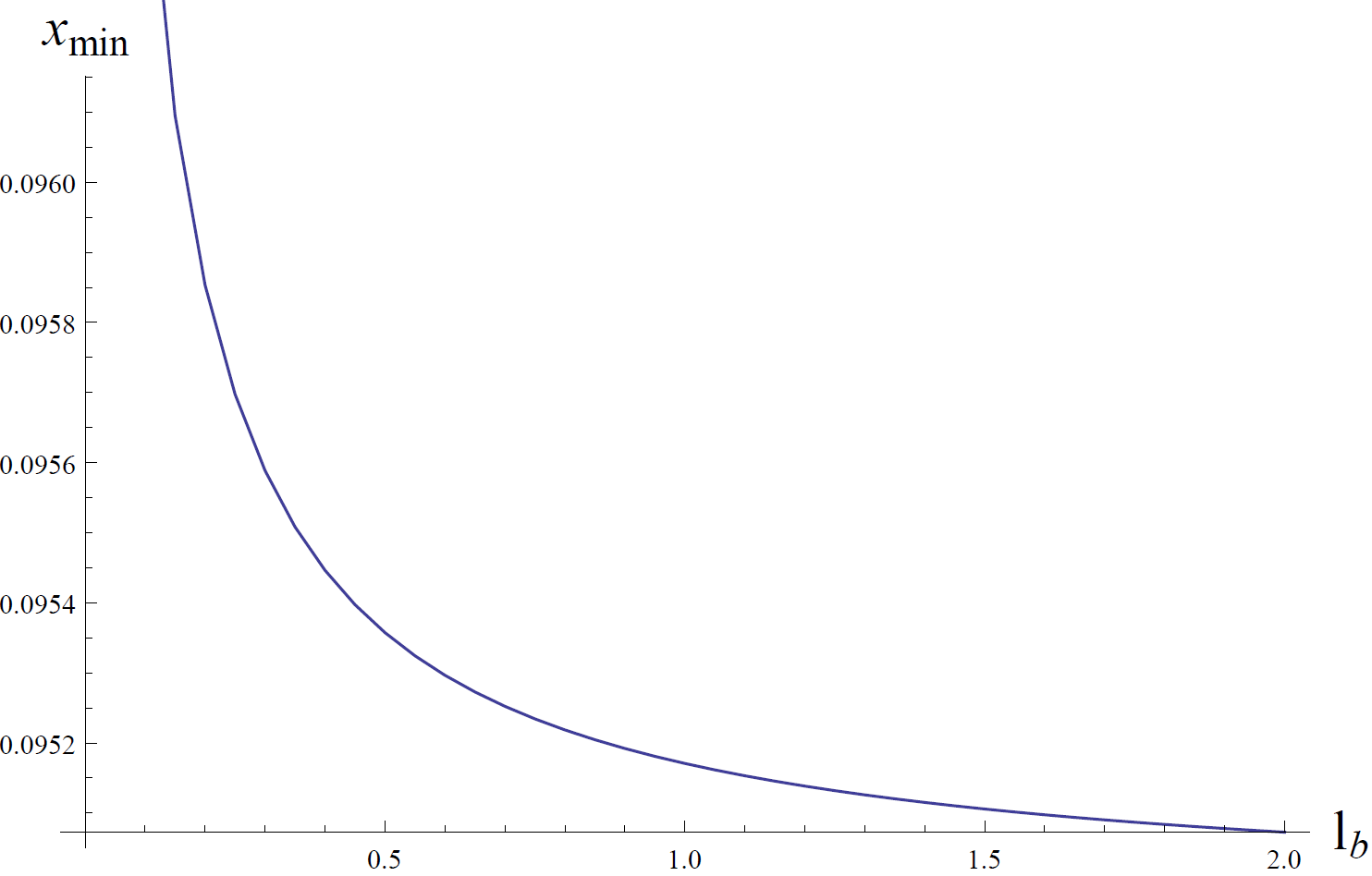}
    \centering
    \caption{x-value of the minimum of the potential for different values of $0<l_{b}<2$ and $\frac{D_{\nu}}{W}=0.14$}
    \label{fig:MinVar}
\end{figure}

\begin{figure}[H]
    \includegraphics[width=7.0cm]{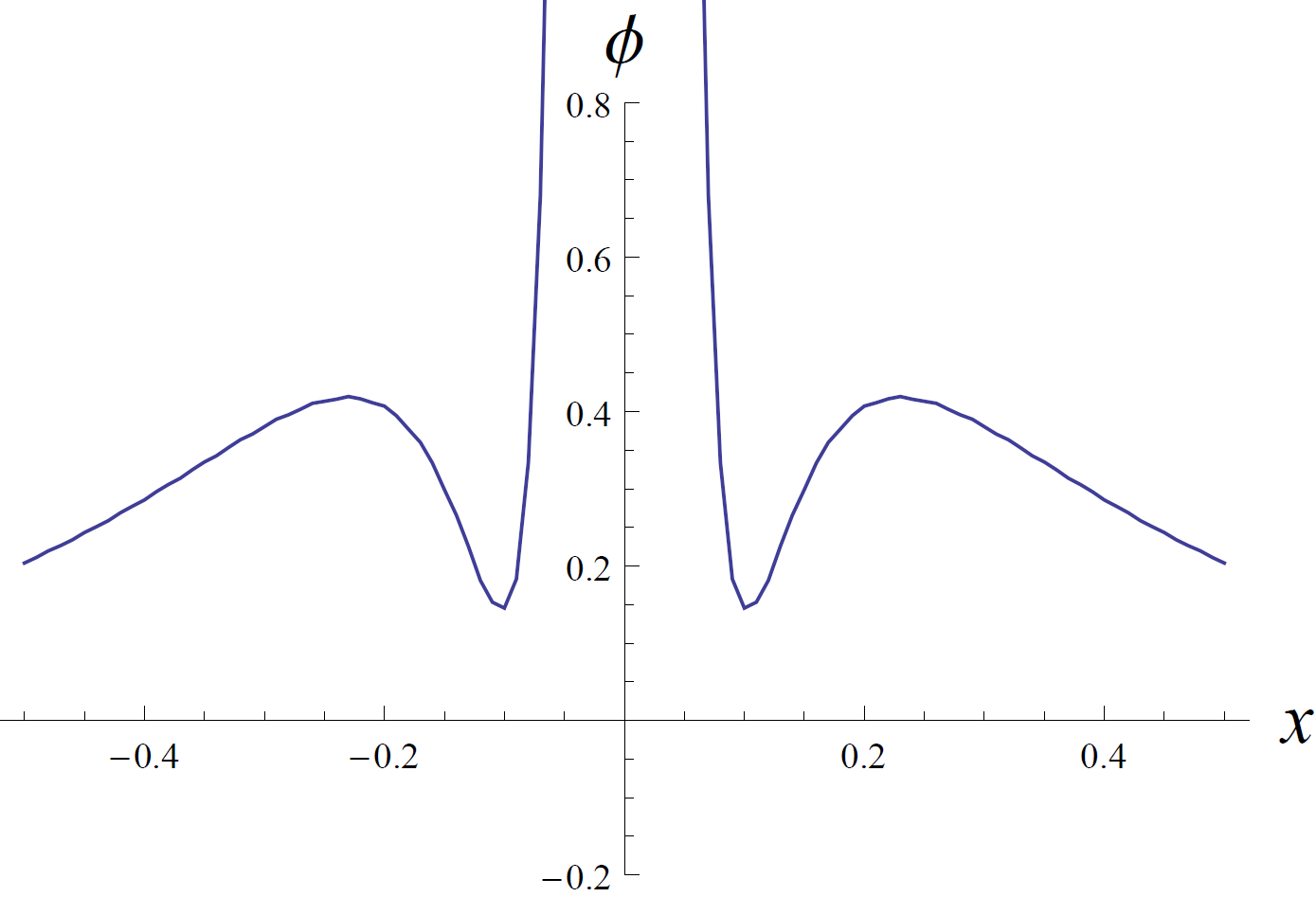}
    \centering
    \caption{$\phi (y=-W/2)$ for no-slip $\frac{l_{b}}{W}=0$ and $\frac{\rho(enW)^{2}}{\eta}=50$ or
$\frac{D_{\nu}}{W}=0.14$}
    \label{fig:FL 0 1}
\end{figure}

\begin{figure}[H]
    \includegraphics[width=7.0cm]{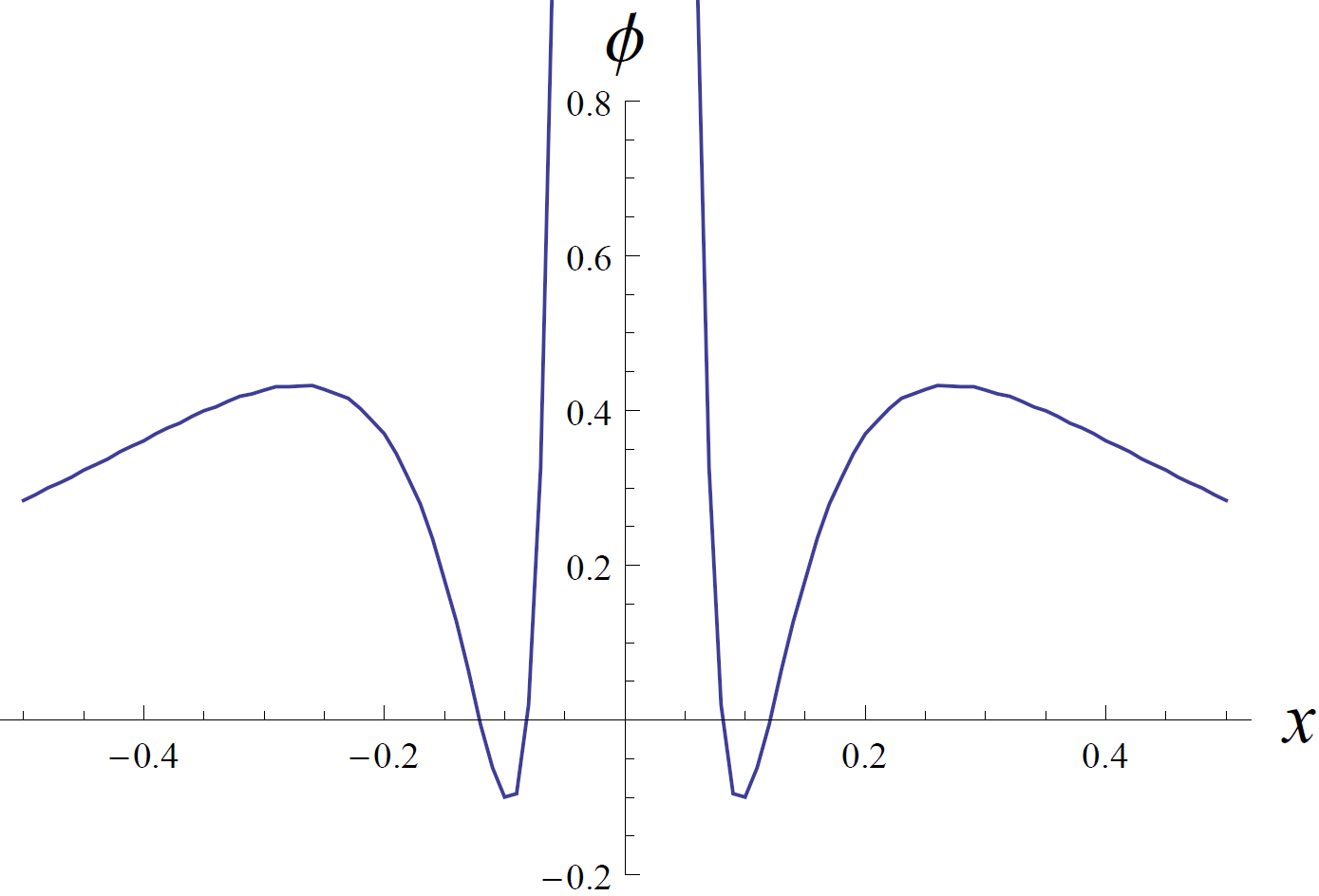}
    \centering
    \caption{$\phi (y=-W/2)$ for partial slip $\frac{l_{b}}{W}=1$ and $\frac{\rho(enW)^{2}}{\eta}=50$
or $\frac{D_{\nu}}{W}=0.14$}
    \label{fig:FL 1 1}
\end{figure}

\begin{figure}[H]
    \includegraphics[width=7.0cm]{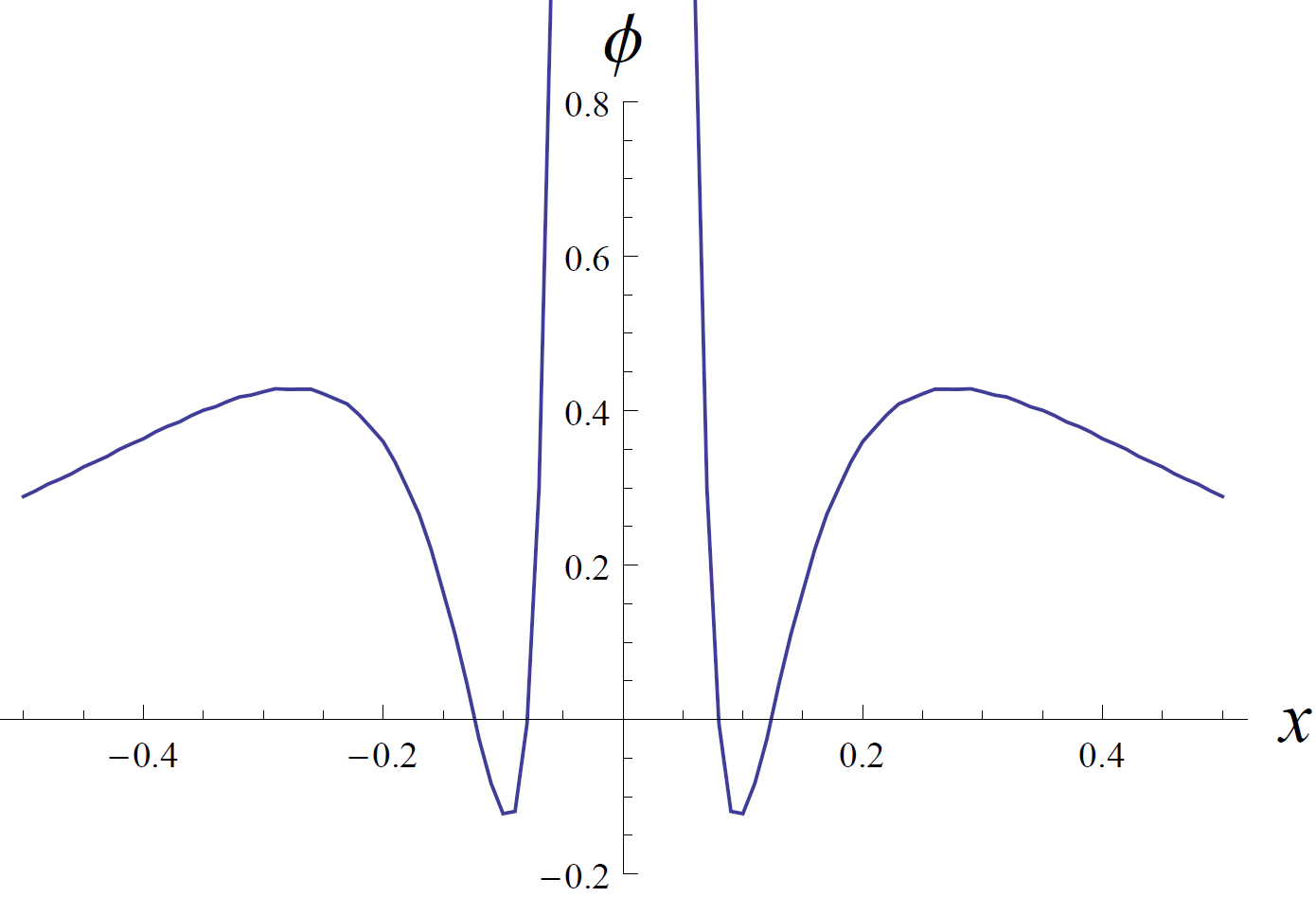}
    \centering
    \caption{$\phi (y=-W/2)$ for free boundary $\frac{l_{b}}{W}=1000$ and $\frac{\rho(enW)^{2}}{\eta}=50$
or $\frac{D_{\nu}}{W}=0.14$}
    \label{fig:FL 1000 1}
\end{figure}

We can now look at the effect that varying the slipping length $l_{b}$ has on the potential for the viscosity corresponding to $\frac{D_{\nu}}{W}=0.14$. From Fig. \ref{fig:MinVar} we see that the minimum of the potential moves closer to the origin for increasing $l_{b}$, hence the vortices move closer to the origin. In addition, from Figs. \ref{fig:FL 0 1} to \ref{fig:FL 1000 1} we see that the minimum value of the potential decreases, i.e. the negative potential becomes more pronounced, when we increase $l_{b}$. This makes sense intuitively, for no-slip conditions $l_{b}=0$ we expect a comparatively weak vortex, since the tangential velocity is zero at the boundary, for free boundary conditions $l_{b}\rightarrow\infty$ this restriction does not exist, so the vortices become more pronounced which manifests itself in a stronger negative potential and the vortices moving closer to the origin; the latter effect being weaker than the other.

\begin{figure}[H]
    \includegraphics[width=1\linewidth]{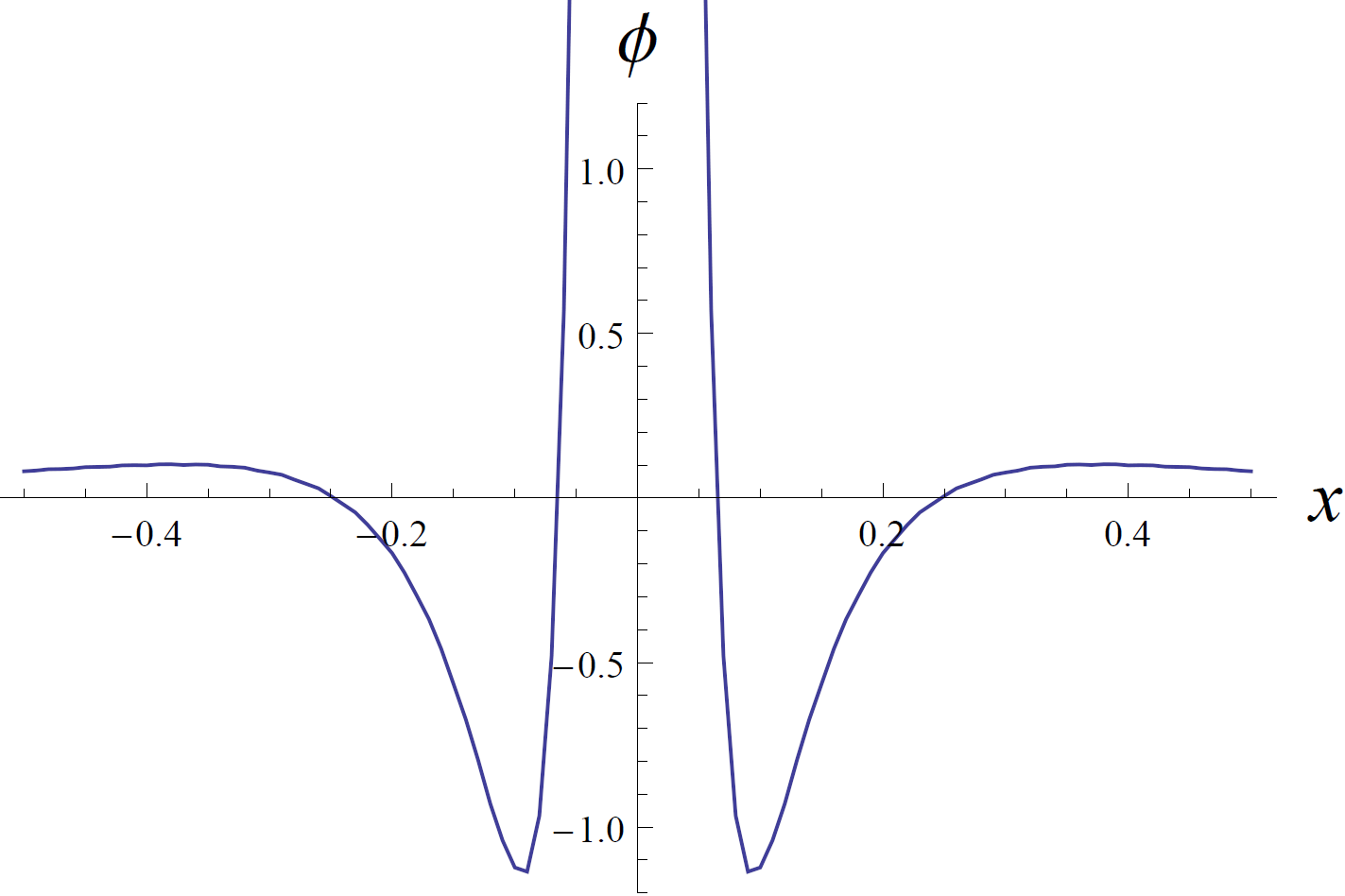}
    \centering
    \caption{$\phi (y=-W/2)$ for no-slip $\frac{l_{b}}{W}=0$ and $\frac{\rho(enW)^{2}}{\eta}=30$
or $\frac{D_{\nu}}{W}=0.18$}
    \label{fig:FL 0 50}
\end{figure}

\begin{figure}[H]
    \includegraphics[width=1\linewidth]{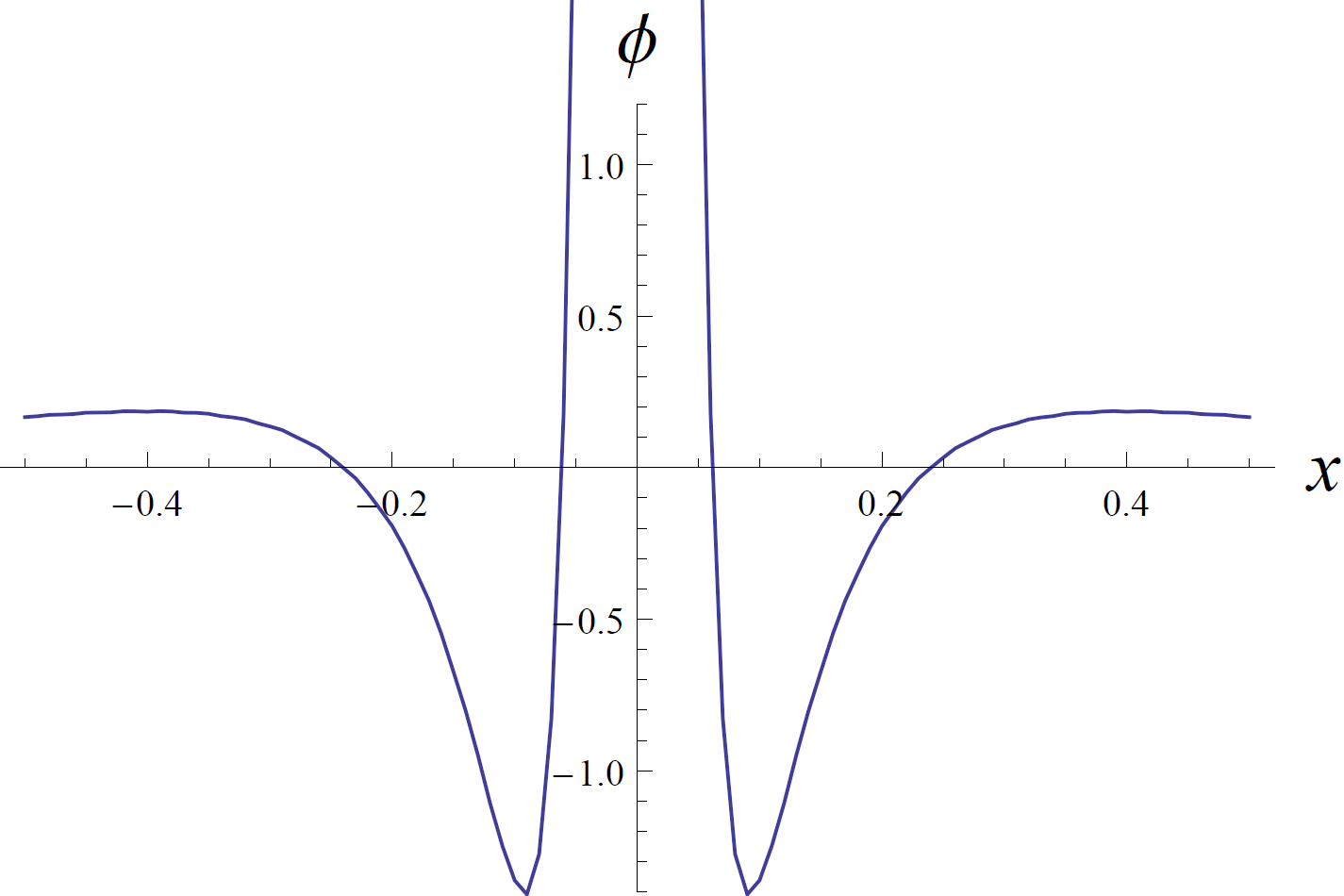}
    \centering
    \caption{$\phi (y=-W/2)$ for partial slip $\frac{l_{b}}{W}=1$and $\frac{\rho(enW)^{2}}{\eta}=30$
or $\frac{D_{\nu}}{W}=0.18$}
    \label{fig:FL 1 50}
\end{figure}

\begin{figure}[H]
    \includegraphics[width=1\linewidth]{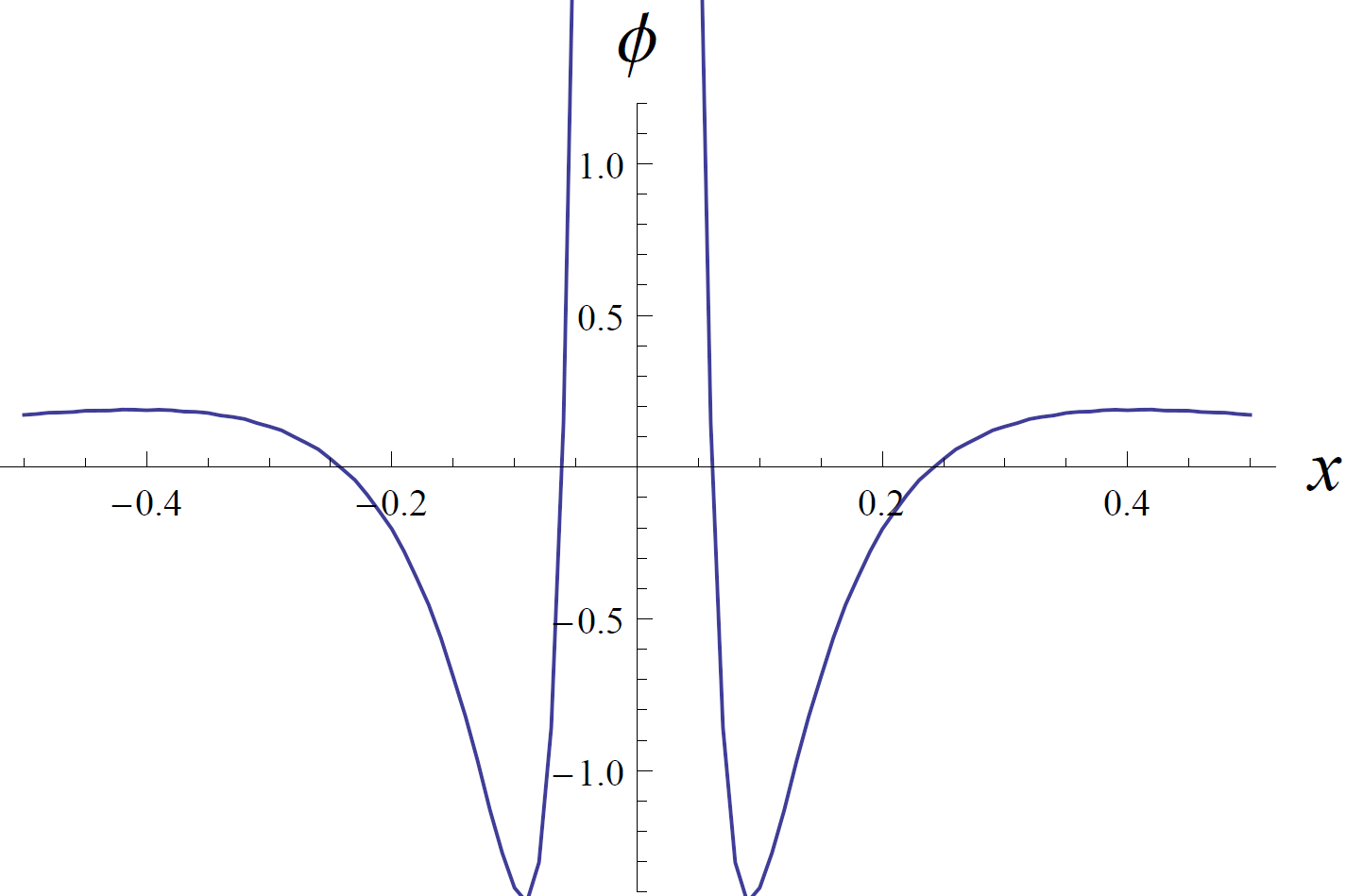}
    \centering
    \caption{$\phi (y=-W/2)$ for free boundary $\frac{l_{b}}{W}=1000$ and $\frac{\rho(enW)^{2}}{\eta}=30$
or $\frac{D_{\nu}}{W}=0.18$}
    \label{fig:FL 1000 50}
\end{figure}

Interestingly, for high enough viscosities as in Figs. \ref{fig:FL 0 50} to \ref{fig:FL 1000 50} the dependence on the slipping length disappears, which is opposite to the non-hydrodynamic regime in which this happens in the previous geometry. Due to high viscosity strong vortices appear regardless of the boundary conditions.
In fact, in contrast to the other geometry, we find three regimes for the dependence of the potential on the boundary conditions. For both high and low viscosities changing $l_{b}$ has little effect.

\begin{figure}
    \includegraphics[width=1\linewidth]{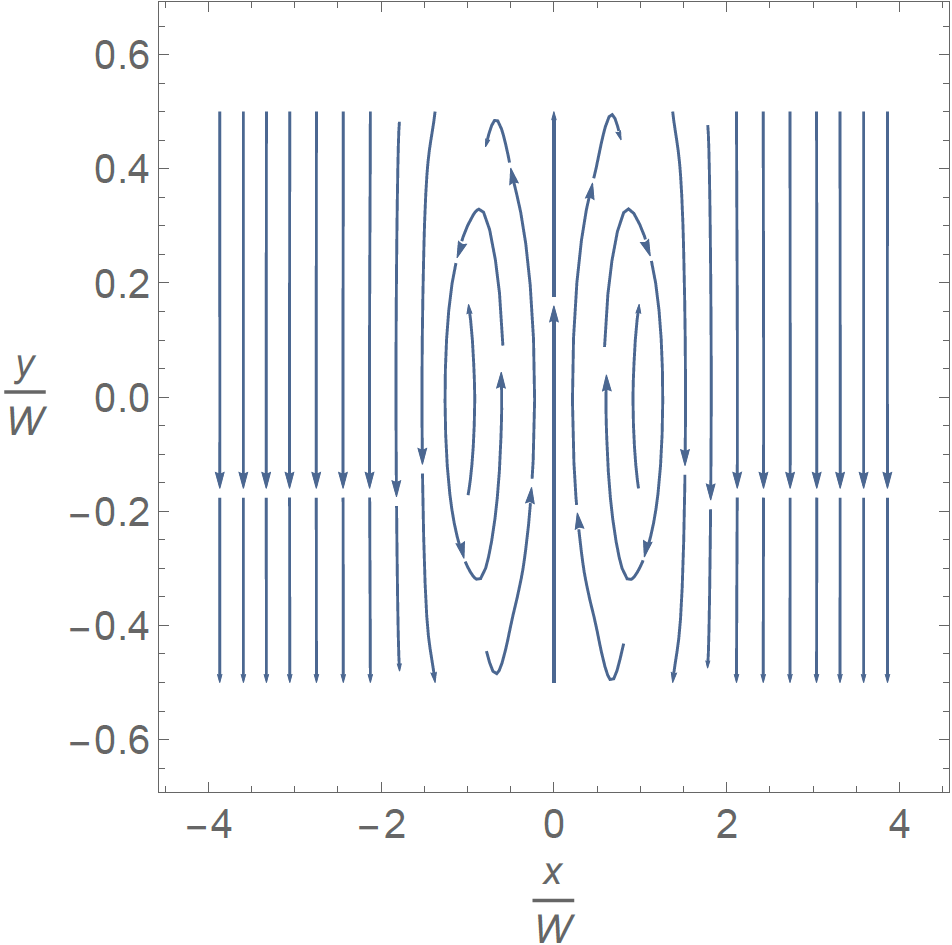}
    \centering
    \caption{Streamlines of the current for partial slip $\frac{l_{b}}{W}=1$ and $\frac{\rho(enW)^{2}}{\eta}=25$ or $\frac{D_{\nu}}{W}=0.2$}
    \label{fig:FLStream}
\end{figure}

In Fig. \ref{fig:FLStream} we plot the streamlines of the current for parameters at which a negative nonlocal resistance appears. In the centre there is a strong flow from the source to the sink. The large vortices on either side have much smaller currents flowing through them. Further away from the centre the flow is in the opposite direction, however it is significantly weaker.

\begin{figure}
    \includegraphics[width=1\linewidth]{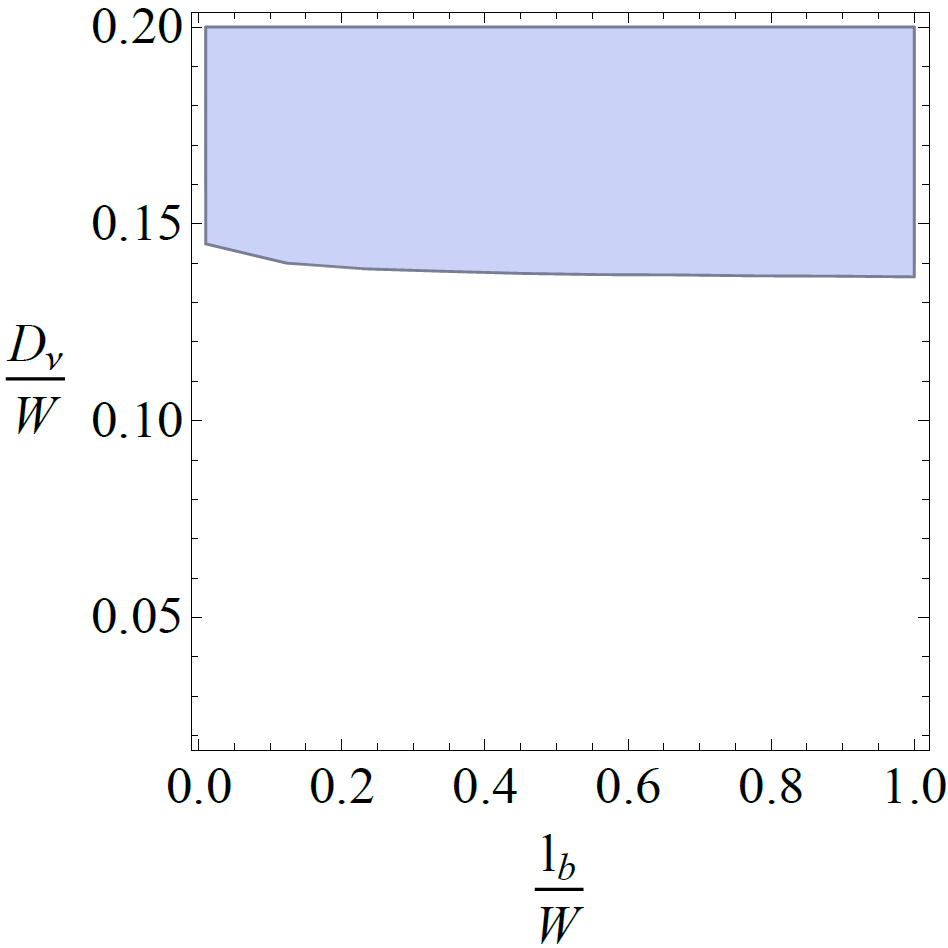}
    \centering
    \caption{coloured region is that region in the $l_{b}$-$D_{\nu}$ parameter space in which $\phi (x,y=-W/2)$ becomes negative (geometry of \cite{FL})}
    \label{fig:FLPhase}
\end{figure}

Again, in Fig. \ref{fig:FLPhase} we plot the region in the $l_{b}$-$D_{\nu}$ parameter space in which $\phi (x,y=-W/2)$ becomes negative, i.e. the region in which we observe a negative nonlocal resistance.

As before, for large enough viscosity, around $D_{\nu}/W>0.15$ there is a negative potential regardless of the boundary conditions. However the dependence on $l_{b}$ is much smaller than in the previous geometry.

\newpage
\subsection{Geometry of \cite{FL}}

\section{Conclusion}

We found that the geometry of \cite{FL} leads to curves of a similar general shape, however it depends on the boundary conditions and the strength of the viscosity whether a negative resistance appears or not. On the other hand in \cite{Geim} the potential can have very different behaviour depending on the two parameters, with the voltage changing sign frequently along the boundary. When conducting experiments, the slip and the viscosity are a priori unknown (although methods exist to theoretically predict the value of the viscosity). The advantage of \cite{Geim} is that the richness of the behaviour of the voltage in the parameter space makes it easier to determine both parameters given the experimental results. If the viscosity is known then the geometry of \cite{FL} would make analysis simpler since the minimum of the potential gives an estimate of which regime of boundary conditions we are in. In fact, if we wanted to determine both the viscosity and the slip, we could use two markers, e.g. the point of the minimum together with the zero.
An advantage of the geometry of \cite{FL} is that it gives a negative nonlocal resistance for smaller values of $\frac{D_{\nu}}{W}$.

Another important question is whether the negative nonlocal resistance is an appropriate identifier for the hydrodynamic regime in graphene as proposed in \cite{FL}. We find that the appearance of a negative nonlocal resistance depends crucially on the size of the viscosity and the slipping length, so this identifier is not totally robust against changes in the boundary conditions. In the case of the geometry of \cite{FL} a more robust signature of the hydrodynamic regime is the presence of a minimum in the potential.  A general trend of the influence of the slipping length on the negative resistance is that the negative resistance is smaller, or unobservable, for the no-slip condition compared to the others. As expected, in both geometries the negative resistance disappears for too small viscosities. However the relevant dimensionless quantity is  $\frac{D_{\nu}}{W}$ so for a small enough width $W$ of the strip, the hydrodynamic regime should be observable.

\paragraph{Acknowledgements}
The author would like to acknowledge many helpful discussions with G. Falkovich and A. Frishman.

\bibliographystyle{ieeetr}
\bibliography{graphene}

\begin{thebibliography}{1}

\bibitem{FluidMech}
G.~Falkovich, {\em
  \href{https://books.google.co.il/books/about/Fluid_Mechanics.html?id=Rt0ZILg5ZcMC&redir_esc=y}{Fluid
  mechanics: A short course for physicists}}.
\newblock Cambridge University Press, 2011.

\bibitem{NoSlip}
M.~A. Day, ``\href{http://link.springer.com/article/10.1007\%2FBF00717588}{The
  no-slip condition of fluid dynamics},'' {\em Erkenntnis}, vol.~33, no.~3,
  pp.~285--296, 1990.

\bibitem{Geim}
I.~{Torre}, A.~{Tomadin}, A.~K. {Geim}, and M.~{Polini},
  ``\href{http://arxiv.org/pdf/1508.00363v1.pdf}{Non-local transport and the
  hydrodynamic shear viscosity in graphene},'' {\em ArXiv e-prints}, Aug. 2015.

\bibitem{FL}
L.~{Levitov} and G.~{Falkovich},
  ``\href{http://arxiv.org/pdf/1508.00836v1.pdf}{Electron Viscosity, Current
  Vortices and Negative Nonlocal Resistance in Graphene},'' {\em ArXiv
  e-prints}, Aug. 2015.

\bibitem{Geim2}
D.~A. {Bandurin}, I.~{Torre}, R.~{Krishna Kumar}, M.~{Ben Shalom},
  A.~{Tomadin}, A.~{Principi}, G.~H. {Auton}, E.~{Khestanova}, K.~S.
  {Novoselov}, I.~V. {Grigorieva}, L.~A. {Ponomarenko}, A.~K. {Geim}, and
  M.~{Polini}, ``\href{http://arxiv.org/abs/1509.04165}{Negative local
  resistance due to viscous electron backflow in graphene},'' {\em ArXiv
  e-prints}, Sept. 2015.

\bibitem{Gurzhi}
R.~Gurzhi,
  ``\href{http://www.turpion.org/php/paper.phtml?journal_id=pu&paper_id=3815}{Hydrodynamic
  effects in solids at low temperature},'' {\em Physics-Uspekhi}, vol.~11,
  no.~2, pp.~255--270, 1968.

\bibitem{PerfFluid}
M.~M{\"u}ller, J.~Schmalian, and L.~Fritz,
  ``\href{http://link.aps.org/doi/10.1103/PhysRevLett.103.025301} {Graphene: A
  nearly perfect fluid},'' {\em Physical review letters}, vol.~103, no.~2,
  p.~025301, 2009.

\end{thebibliography}

\hspace{20 mm}
\newpage

\pagebreak
\section{Appendix}
\subsection{Calculation of $\phi$}
Due to the symmetry of the geometry in \cite{FL} this problem can be solved by hand. We expect the potential to be symmetric with respect to $x$ and anti-symmetric with respect to $y$. Hence the Fourier transform must satisfy

\begin{equation}
\hat{\phi}(k,y)=\hat{\phi}(-k,y)
\end{equation}
and

\begin{equation}
\hat{\phi}(k,y)=-\hat{\phi}(k,-y)
\end{equation}
It follows that the coefficients from (\ref{solution}) satisfy $a{}_{1}=-a_{2}$ and $a{}_{3}=a_{4}$. We then find, using the Fourier transformed boundary conditions, that

\begin{equation}
\hat{\phi}(k,y)=-\frac{e}{\sigma_{0}}\frac{a{}_{1}(k)\sinh(ky)}{k^{2}}
\end{equation}
where

\small
\begin{equation}
a{}_{1}(k)=\frac{I}{e}\frac{-ke^{-\mid k\mid l}}{2\cosh(kW/2)+\frac{2(kD_{\nu})^{2}}{1+(kD_{\nu})^{2}}\cosh(k \tilde{W}/2)\mathrm{f}(k)}
\end{equation}

\begin{equation}
\mathrm{f}(k)=\frac{4\cosh(kW/2)+\frac{2}{kl_{b}}\sinh(kW/2)}{\frac{2}{kl_{b}}\frac{\mid k\mid D_{\nu}}{\sqrt{1+(kD_{\nu})^{2}}}\sinh(k \tilde{W}/2)+2 \frac{1+2(kD_{\nu})^{2}}{1+(kD_{\nu})^{2}}\cosh(k \tilde{W}/2)}
\end{equation}
\normalsize
where
\begin{equation}
\tilde{W}=W\sqrt{1+\frac{1}{(kD_{\nu})^{2}}}
\end{equation}

We must then Fourier transform to obtain $\phi$, this is done numerically.

In the geometry of \cite{Geim} this symmetry argument does not work and the algebra is more tedious, hence we employ Mathematica to solve the problem.

\end{document}